# DRL²FC: An Attack-Resilient Controller for Automatic Generation Control Based on Deep Reinforcement Learning

Vasileios Dimitropoulos [1*†], Andreas D. Syrmakesis [1**†], Nikos D. Hatziargyriou [1***]

[1] School of Electrical & Computer Engineering, National Technical University of Athens, 9 Iroon Polytechniou str, 15772 Zografou, Greece

†: Equal contribution

*: el07225@mail.ntua.gr, **: Corresponding author, asirmakesis@power.ece.ntua.gr, ***: nh@power.ece.ntua.gr

***Abstract*** - Power grids heavily rely on Automatic Generation Control (AGC) systems to maintain grid stability by balancing generation and demand. However, the increasing digitization and interconnection of power grid infrastructure expose AGC systems to new vulnerabilities, particularly from cyberattacks such as false data injection attacks (FDIAs). These attacks aim at manipulating sensor measurements and control signals by injecting tampered data into the communication mediums. As such, it is necessary to develop innovative approaches that enhance the resilience of AGC systems. This paper addresses this challenge by exploring the potential of deep reinforcement learning (DRL) to enhancing the resilience of AGC systems against FDIAs. To this end, a DRL-based controller is proposed that dynamically adjusts generator setpoints in response to both load fluctuations and potential cyber threats. The controller learns these optimal control policies by interacting with a simulated power system environment that incorporates the AGC dynamics under cyberattacks. The extensive experiments on test power systems subjected to various FDIAs demonstrate the effectiveness of the presented approach in mitigating the impact of cyberattacks.

## 1. Introduction

In recent years, Smart Grids (SGs) have revolutionized the way power systems operate, by leveraging information and communication technologies (ICT) to enhance their monitoring, control, and communications [1, 2]. However, the high coupling of SGs with ICT has made them susceptible to cyberattacks. A critical power system automation, that is directly is exposed to digital risks due to its reliance on ICT, is the Automatic Generation Control (AGC) [3, 4]. The role of this mechanism is to maintain the energy balance in electrical systems. An emerging cyber threat against these automation systems are the False Data Injection Attacks (FDIAs). These types of cyberattacks inject tampered data into the sensor measurements and control signals of AGC to manipulate control decisions and compromise grid stability [5, 6].

The importance of AGC has prompted researchers to propose cybersecurity measures to protect them against malicious activities. An interesting approach towards cyberattack detection and mitigation in AGC is the reinforcement learning (RL). RL is a category of machine learning algorithms where an agent learns to make decisions by interacting with an environment, receiving feedback in the form of rewards or penalties, and adjusting its actions to maximize cumulative rewards over time. More specifically, a deep Q network-based optimal control and an RL feedback control-based compensation are proposed in [7] for AGC against cyber threats. However, these strategies mainly focus on load altering attacks and cannot be scaled to other measurement or control signal attacks. From the perspective of the attacker, the problem of constructing undetectable attacks is investigated in [8] as a multi-object partially observable Markov decision process, along with a flexible reward function that maximizes the impact of the RL-based attack. While this method offers valuable insights regarding the patterns of the attackers, it does not provide an analytical RL method of eliminating them. In [9], RL is used to generate a dataset that simulates AGC under normal and attack conditions. Then, this dataset is used to train a Long Short-Term Memory (LSTM)-based attack detector that can identify cyberattacks but cannot tackle them.

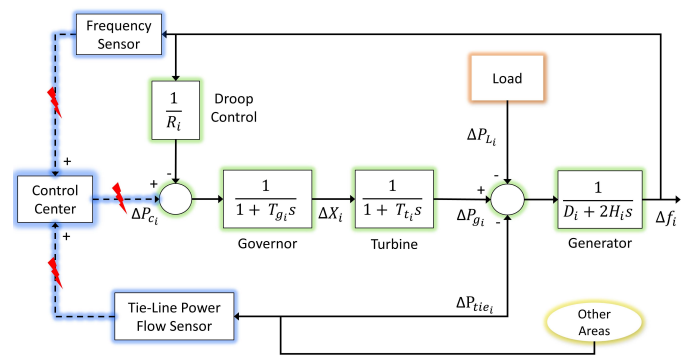

***Fig. 1.*** *Block diagram of the $i^{th}$ AGC area*

The limitations of the existing RL-based solutions towards the cyber resilience strengthening of AGC against attacks have inspired the development of the proposed work. Particularly, the introduced paper addresses these issues by leveraging Deep Reinforcement Learning (DRL) to develop a controller, called DRL²FC, that mitigates the impact of various types of FDIAs. The main contributions of this paper are summarized as:

- The implementation of a novel, attack-resilient, DRL-based controller for AGC.
- The capability of DRL²FC to distinguish FDIAs from other external disturbances.
- The scalability of DRL²FC to various power systems through experimental validation.

## 2. System Modelling

The purpose of the AGC is to maintain the energy equilibrium between generation and demand within a power grid. A proper indicator for this balance is the deviations of frequency and interarea tie-line power flow measurements from their nominal values. The AGC receives these measurements, applies its primary and secondary control algorithms and then computes the control signals to adjust the generation according to the load changes. This functionality is formally described by the differential-algebraic equations of AGC in Fig. 1, which illustrates the block diagram of its $i^{th}$ area. Particularly, the primary control is responsible for stabilizing the frequency after a load disturbance and is implemented locally on the generators through their governors (green blocks in Fig. 1). The secondary control eliminates the remaining steady-state error caused by the primary control and restores the frequency and tie-line power interchanges to their nominal values. This type of control is applied using telemetries and remote controls based on ICT (blue blocks in Fig. 1). Therefore, AGC is inevitably vulnerable to FDIAs against the network-transmitted data as indicated by the red points in Fig.1.

## 3. Proposed Attack-resilient Control

In this work, an attack-resilient controller, called DRL$^2$FC, is developed and trained for strengthening the cyber resilience of AGC systems. The controller is implemented as an agent using a Deep Q-Network (DQN) architecture. The agent receives AGC measurements, i.e. $s := (\Delta f_i, \Delta P_{tie_i})$, acts on the AGC environment and receives rewards based on its actions. The action space is defined as the concatenation of the generation control commands, $a := (P_{C_i})$.

The decision-making process of the agent follows an ε-greedy policy to explore the state-action space. Mathematically, this policy can be expressed as:

$$\pi(a|s) = \begin{cases} 1 - \varepsilon + \frac{\varepsilon}{|a(s)|} & \text{if } a = argmax_a Q(s,a) \\ \frac{\varepsilon}{|a(s)|} & \text{otherwise} \end{cases} \quad (1),$$

where $\pi(a|s)$ represents the probability of selecting action $a$ given state $s$, $Q(s,a)$ denotes the quality-value function, $|a(s)|$ is the cardinality of the action space at state $s$, and $\varepsilon$ denotes the exploration probability.

The policy is optimized through stochastic gradient descent (SGD) using a replay memory mechanism. The optimization process seeks to minimize the temporal difference error between the target Q-value function, $Q'(s',a')$, and the current estimate, $Q(s,a)$, which can be formulated as:

$$\delta = r + \gamma max_{a'} Q'(s',a') - Q(s,a) \quad (2),$$

where $\delta$ represents the temporal difference error, $\gamma$ is the discount factor, $s'$ is the next state, and $r$ denotes the immediate reward, which is determined by:

$$R(s,a,s') = -\Delta t \sum_i \left( (\beta_i \Delta f_i(t))^2 + \left( \sum_j \Delta P_{tie_{ij}} \right)^2 \right) \quad (3),$$

where $\beta_i$ is the frequency bias factor. Therefore, the agent is penalized until it restores the frequency and tie-line power flow to their nominal values after disturbances.

The parameters of the DQN are updated iteratively to minimize the loss function, which is defined as the mean squared error between the target Q-value and the predicted Q-value. The training process can be described as:

$$Loss = E_{s,a,r,s'}[(r + \gamma max_{a'} Q'(s',a') - Q(s,a))^2] \quad (4),$$

where the expectation is taken over samples from the replay memory.

## 4. Experimental Results

In this Section, a comparative analysis between DRL$^2$FC and other, conventional AGC techniques (fine-tuned PID, LQR, MPC) is conducted to demonstrate the effectiveness of the proposed methodology. The compared controllers are evaluated on a two-area AGC system, where the area structure is shown in Fig. 1 and the area parameters are adopted by [6]. The simulated events are the following: (a) a 0.01 p.u. step disturbance in the load of area 1 at t=5s, followed by step FDIA signal of 0.01 p.u. launched against the frequency sensor of area 2, (b) a coordinated pulse FDIA on tie-line power flow sensors of both areas, and (c) a simultaneous load disturbance of 0.02 p.u. and a ramp FDIA on area 1 control signal. The experimental results are illustrated in Figure 2: when load changes occur, the performance of the DRL$^2$FC is comparable to the standard frequency control methods. However, conventional controllers fail to maintain zero frequency deviation under FDIAs, while the proposed DRL-based AGC effectively restores system frequency to its nominal value. Therefore, it numerically validated that the DRL$^2$FC enhances the cyber resilience of AGC and outperforms other, standard power system frequency controllers.

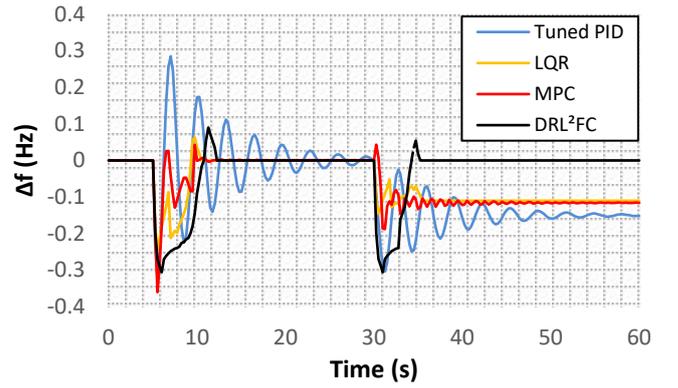

***Fig. 2.*** *Frequency deviation under different controllers – scenario (a)*